\documentclass[conference]{IEEEtran}
\IEEEoverridecommandlockouts
\usepackage[noadjust]{cite}
\usepackage{amsmath, amssymb, amsfonts}
\usepackage{algorithmic}
\usepackage{graphicx}
\usepackage{textcomp}
\usepackage{xcolor}
\usepackage{listings}
\usepackage[hyphens]{url}
\usepackage{booktabs}
\usepackage{multirow}

\usepackage{tikz}
\usetikzlibrary{shapes.geometric, arrows, positioning, calc}

\def\BibTeX{{\rm B\kern-.05em{\sc i\kern-.025em b}\kern-.08em
    T\kern-.1667em\lower.7ex\hbox{E}\kern-.125emX}}

\begin{document}

\title{Automated Non-Functional Requirements Generation in Software Engineering with \\Large Language Models: A Comparative Study}

\author{
\IEEEauthorblockN{Jomar Thomas Almonte}
\IEEEauthorblockA{\textit{Engineering, University Park}\\
The Pennsylvania State University\\
Pennsylvania, USA\\
jzt5803@psu.edu}
\and
\IEEEauthorblockN{Santhosh Anitha Boominathan}
\IEEEauthorblockA{\textit{Engineering, Great Valley}\\
The Pennsylvania State University\\
Pennsylvania, USA\\
sfa5971@psu.edu}
\and
\IEEEauthorblockN{Nathalia Nascimento}
\IEEEauthorblockA{\textit{Engineering, Great Valley}\\
The Pennsylvania State University\\
Pennsylvania, USA\\
nnascimento@psu.edu}
}

\maketitle

\begin{abstract}
Neglecting non-functional requirements (NFRs) early in software development can lead to critical challenges. Despite their importance, NFRs are often overlooked or difficult to identify, impacting overall software quality. To support requirements engineers in eliciting NFRs, we developed a framework that leverages Large Language Models (LLMs) to derive quality-driven NFRs from functional requirements (FRs). Using a custom prompting technique within a Deno-based pipeline, the system identifies relevant quality attributes for each functional requirement and generates corresponding NFRs, aiding systematic integration.

A crucial aspect is evaluating the quality and suitability of these generated requirements. Can LLMs produce high-quality NFR suggestions? 
Using 34 functional requirements—selected as a representative subset of 3,964 FRs—the LLMs inferred applicable attributes based on the ISO/IEC 25010:2023 standard, generating 1,593 NFRs. A horizontal evaluation covered three dimensions: NFR validity, applicability of quality attributes, and classification precision. Ten industry software quality evaluators, averaging 13 years of experience, assessed a subset for relevance and quality. The evaluation showed strong alignment between LLM-generated NFRs and expert assessments, with median validity and applicability scores of 5.0 (means: 4.63 and 4.59, respectively) on a 1–5 scale. In the classification task, 80.4\% of LLM-assigned attributes matched expert choices, with 8.3\% near misses and 11.3\% mismatches. A comparative analysis of eight LLMs highlighted variations in performance, with \texttt{gemini-1.5-pro} exhibiting the highest attribute accuracy, while \texttt{llama-3.3-70B} achieved slightly higher validity and applicability scores.
These findings provide insights into the feasibility of using LLMs for automated NFR generation and lay the foundation for further exploration of AI-assisted requirements engineering.


\end{abstract}

\begin{IEEEkeywords}
Non-Functional Requirements, Large Language Models, Requirements Elicitation, ISO/IEC 25010, Automated Requirements Engineering.
\end{IEEEkeywords}

\section{Introduction}
The elicitation of requirements is a fundamental phase in the software development lifecycle, as it directly influences the quality, maintainability, and success of the final product. A complete specification should include both functional requirements (FRs), which define the different functionalities the system should perform, and non-functional requirements (NFRs), which define attributes of how the system should perform these functionalities \cite{mizouni2010towards}. While FRs describe what a system must do, NFRs—such as performance, security, and usability—often end up distinguishing competing products and are equally critical \cite{LaplanteKassab}. Despite their importance, NFRs are often ambiguously defined or overlooked, leading to costly rework and compromised system quality.

Many FRs inherently imply specific NFRs \cite{li2015stakeholder} or provide context for them, but recognizing these relationships often relies on the expertise of requirements engineers. For instance, a functional requirement specifying a ``Login" feature suggests the need for security-related NFRs. However, ensuring comprehensive NFR coverage can be challenging, even for experienced engineers. Overlooking critical NFRs can introduce significant risks, while addressing irrelevant or inapplicable attributes may lead to unnecessary costs. This highlights the need for a systematic approach to identifying the most relevant quality attributes for each FR. Our approach leverages large language models (LLMs) to analyze FRs, infer applicable quality attributes, and generate corresponding NFRs, providing structured guidance to enhance NFR coverage and support a more comprehensive specification process.

Recent advancements in natural language processing (NLP) and the advent of LLMs have enabled automation in complex language understanding tasks \cite{Brown2020}. Requirements engineering has also begun integrating LLMs \cite{cheng2024generative,porter2025requirements,lano2024introduction}. Although prior research has applied AI techniques to the identification and classification of NFRs, as well as distinguishing NFRs from FRs and categorizing them for better prioritization \cite{8559686,10181313,li2015stakeholder}, to the best of our knowledge, no previous work has explored the use of LLMs to generate NFRs. 

In this study, we investigate the feasibility of using LLMs to generate NFRs from FRs, evaluating their ability to systematically infer quality attributes and produce suitable NFRs. Using 34 functional requirements, selected as a representative subset from a dataset of 3,964 FRs, LLMs identified applicable quality attributes based on the ISO/IEC 25010:2023 quality model and generated 1,593 NFRs. To assess the quality of these NFRs, we conducted a dual-evaluation strategy with 10 expert evaluators—experienced professionals averaging 13 years in roles such as Senior Software Engineer, Principal Program Manager, and Software Quality Engineer. This evaluation comprised two tasks: an ordinal classification to score the generated NFRs and an attribute selection process to validate their assigned quality attributes.

\textbf{Our contributions are multifold:}
\begin{enumerate}
    \item Development of an automated system for NFR generation – Implemented a systematic approach using LLMs to derive NFRs from FRs, integrating a custom prompting pipeline aligned with ISO/IEC 25010:2023 standards.
    \item Design of an optimized prompting strategy – Developed a structured multi-technique prompting approach, enhancing generation precision through role descriptions, constraint enforcement, and in-context learning.
    \item Expert-driven validation of generated NFRs – Conducted a structured evaluation with domain experts to assess the generated NFRs.
    \item Comparative evaluation of LLMs for NFR generation – Assessed multiple LLMs to determine their effectiveness in generating relevant NFRs.
    \item Creation of a specialized and open dataset  \cite{HumanEvaluationDataset} – Curated a labeled dataset comprising FRs, corresponding NFRs, and expert evaluations, supporting benchmarking and further research.
    \item Establishment of a research agenda – Outlined future directions for advancing AI-assisted NFR generation.
\end{enumerate}

This paper is organized as follows. Section \ref{sec:related} reviews related work. Section \ref{sec:rqs} outlines the research questions and objectives. Sections \ref{sec:methodology} and \ref{sec:evaluation} describe the methodology, experiment design, and evaluation approach. Sections \ref{sec:conclusion} and \ref{sec:discussion} present the experimental results and discussion. Finally, Sections \ref{sec:conclusion} and \ref{sec:future} provide concluding remarks and a research agenda for future work. 


\section{Background and Related Work} \label{sec:related}

\subsection{Non-Functional Requirements (NFRs)}
NFRs encompass a system’s quality-related attributes (e.g. reliability, usability, performance) as well as constraints (e.g. legal or design constraints) under which the system must operate \cite{glinz2007non}. There are different classification schemes \cite{Chung2009}, but they categorize NFRs according to the specific quality concerns they address. These requirements include, but are not limited to, performance efficiency, security, usability, and maintainability.

Different methodologies exist for identifying NFRs in a system project, including defining them based on business goals, eliciting them from stakeholder needs, and deriving them from functional requirements \cite{paech2004non}. However, establishing clear relationships between non-functional requirements (NFRs) and their associated functional requirements can be challenging, as these connections are often implicit and difficult to structure \cite{li2015stakeholder}. As Li et al. show, a unified modeling approach is crucial so that NFRs are not treated entirely separate from their functional counterparts \cite{li2015stakeholder}.

\subsection{Large Language Models (LLMs)}
Large language models (LLMs) based on the Transformer architecture, such as GPT, have demonstrated impressive capabilities in generating coherent and contextually relevant text, capturing long-range dependencies \cite{Brown2020}. Recent studies on prompt-based approaches (e.g. \cite{Schick2021}) indicate that incorporating explicit role descriptions and examples in context can further improve task performance, such as requirement classification.

\subsection{ISO/IEC 25010:2023 Standard}
The ISO/IEC 25010:2023 standard \cite{ISO25010} provides a structured framework for evaluating software quality, defining nine key characteristics: \emph{functional suitability, performance efficiency, compatibility, usability, reliability, security, maintainability, flexibility,} and \emph{safety}. These characteristics, further broken down into subcharacteristics, serve as a reference for specifying, measuring, and assessing software quality throughout the product lifecycle. The standard is widely used by developers, acquirers, quality assurance professionals, and independent evaluators to ensure software meets both functional and non-functional requirements across different operational contexts.

\subsection{Related Work}
Requirements Engineering has long been recognized as a promising area for the application of AI techniques, including LLMs \cite{umar2024advances}\cite{cheng2024generative}. In a recent comprehensive review, Umar and Lano \cite{umar2024advances} analyzed 85 papers discussing automated RE from a range of methodological perspectives. Notably, among these studies, only three \cite{cleland2007automated,casamayor2010identification,haque2019non} specifically addressed non-functional requirements (NFRs). These works explored, respectively, automated classification of NFRs \cite{cleland2007automated}, semi-supervised approaches for identifying NFRs in textual specifications \cite{casamayor2010identification}, and methods for classifying NFRs \cite{haque2019non}.

More recently, Cheng et al.\cite{cheng2024generative} investigated the use of generative AI for various RE tasks, citing Rejithkumar et al.\cite{Rejithkumar2025NICE} as one of the few examples where LLMs were leveraged to verify NFRs in a specific application domain. Despite these advances, the broader integration of LLMs into NFR elicitation and classification remains underexplored. This gap is further underscored by Damirchi et al.\cite{damirchi2023non}, who reviewed techniques for extracting NFRs in AI-based systems and concluded that no reliable, standardized solution has yet emerged. Their study also highlighted new and increasingly important NFRs for AI applications, such as Explainability and Transparency.

Identifying and classifying NFRs is a critical task in RE, as accurate classification helps ensure proper allocation, prioritization, and management of requirements \cite{8559686}. For instance, Khan et al.\cite{10181313} proposed using transfer learning to identify and classify NFRs into 12 categories, though they did not align their model with any standardized quality framework. Other relevant research has focused on using AI to transform informal requirements into formal specifications \cite{li2015stakeholder}, as well as on employing LLMs to assist in formalizing requirements \cite{lano2024introduction}. More recently, Porter et al.\cite{porter2025requirements} explored how LLMs might automatically evaluate requirement quality. 

While these studies demonstrate a growing interest in AI-driven approaches for NFR-related tasks, they also highlight a significant gap in robust, automated solutions for NFR elicitation. Existing research predominantly focuses on classification, formalization, or quality assessment, rather than on directly generating NFRs. To the best of our knowledge, this paper is the first to examine how LLMs can generate NFRs by systematically deriving them from functional requirements. Our contributions include the evaluation of multiple LLMs for automated NFR generation, assessment of their effectiveness in producing high-quality NFRs aligned with ISO/IEC 25010:2023, and the development of a labeled dataset encompassing both functional and non-functional requirements. 

\section{Research Objectives and Questions} \label{sec:rqs}
The primary objectives of this research are:
\begin{itemize}
    \item Evaluate the effectiveness of different LLMs in generating NFRs from Functional Requirements.
    \item Develop an automated pipeline that implements a custom prompting technique and adheres to the ISO/IEC 25010:2023 standards.
    \item Validate the outputs generated through human evaluators by systematically sampling the responses.
\end{itemize}

Consequently, we address the following research questions:
\begin{itemize}
    \item \textbf{RQ1:} How effective are LLMs in generating non-functional requirements that are clear, relevant, and comprehensive?
    \item \textbf{RQ2:} Among the eight LLMs tested, which model delivers the most reliable NFR generation according to the ISO/IEC 25010:2023 criteria?
    \item \textbf{RQ3:} Can a custom prompting technique, enhanced with role descriptions and in-context learning, improve the precision of NFR generation compared to baseline methods?
\end{itemize}

\section{Methodology} \label{sec:methodology}
\subsection{Data Collection and Dataset Description}
\label{sec:data_collection}
For this study, we use functional requirements sourced from \textbf{FR\_NFR\_dataset} \cite{FR_NFR_Dataset}. This dataset, extracted in part from the PURE dataset \cite{PURE_Dataset}, supports the requirements engineering phase by providing a collection of software requirements expressed in natural language. Key characteristics of the dataset include:
\begin{itemize}
    \item A total of 6,118 requirements, with 3,964 classified as functional requirements (FR) and 2,154 as non-functional requirements (NFR).
    \item Each entry consists of a requirement text description and a category label that defines its nature (functional or non-functional).
    \item The dataset was manually curated by sifting through Software Requirements Specifications (SRS) documents and annotating requirements accordingly.
    \item Licensed under CC BY 4.0, the dataset is intended to support research in requirements engineering, natural language processing, and machine learning.
\end{itemize}

To simulate a realistic scenario for a small-to-midsize software system, we selected a subset of 34 functional requirements (FR) from the dataset. This number represents the average FR count observed in a corpus of 15 SRS documents, which is a subset of the PURE dataset with 79 SRS documents. From these 79 SRS documents, we selected 15 using the following criteria, partially adapted from Asif et al. \cite{Asif2019} and tailored to our use case:

\begin{itemize}
    \item Documents must be encoded between 2008 and 2011.
    \item SRS should be in a structured or semi-structured format.
    \item SRS should define a software system.
\end{itemize}

The resulting subset of FRs was drawn from the following 15 SRS documents, detailed in Table~\ref{tab:srs_subset}. These documents were chosen to ensure a diverse representation of software systems, including web-based and mobile-based applications. The selected FRs will be stratified into categories such as user authentication, data processing, interface interactions, and system integration to ensure balanced coverage of system functionalities. While we aimed for a sample size representative of mid-sized systems (34 FRs), we acknowledge that a larger dataset would allow for greater statistical power in future studies.

\begin{table}[hbt!]
    \centering
    \caption{Subset of SRS Documents and Functional Requirements}
    \label{tab:srs_subset}
    \begin{tabular}{p{0.3cm} p{1.3cm} p{4cm} p{0.7cm} p{0.5cm}}
        \toprule
        \textbf{No.} & \textbf{Project Name} & \textbf{Topic} & \textbf{Year} & \textbf{FRs} \\
        \midrule
        1 & Email & Statewide enterprise e-mail system & 2009 & 95 \\
        2 & GParted & Partition editor GUI & 2010 & 24 \\
        3 & opensg 0.1 & Information management framework & 2011 & 18 \\
        4 & Split merge & PDF manipulation & 2010 & 13 \\
        5 & Fishing Logbook & Electronic fishing vessel logbook & 2010 & 35 \\
        6 & home 1.3 & Digital home system & 2010 & 33 \\
        7 & Gaia & Catalog data retrieval & 2009 & 27 \\
        8 & warc III & Archive file manipulation & 2009 & 35 \\
        9 & Library System & System administration for an integrated library system & 2009 & 51 \\
        10 & Peppol & General purpose file and archive manager application & 2009 & 11 \\
        11 & VUB & Publication management system & 2008 & 31 \\
        12 & Video Search & Search video in multiple search engines & 2009 & 16 \\
        13 & Caiso & Black Start Capability Plan & 2008 & 54 \\
        14 & KeePass & Password Safe & 2008 & 33 \\
        15 & Peering & Internetworking of Content Delivery Network & 2008 & 27 \\
        \bottomrule
    \end{tabular}
\end{table}

From these 503 FRs across the 15 documents, we selected a subset of 34 FRs to match the average number of FRs typical for mid-size systems.

\subsection{LLM Selection and Configuration}
\label{sec:llm_selection}
This study benchmarks eight LLMs from multiple providers listed in Table~\ref{tab:llm_details}. These models were selected from the top leaderboards based on their comparable performance scores. All models receive standardized prompts to ensure a fair and consistent comparison.

\begin{table}[ht]
    \caption{LLM Configuration}
    \label{tab:llm_details}
    \centering
    \begin{tabular}{|l|c|c|c|}
        \hline
        \textbf{Provider} & \textbf{Model} & \textbf{Temperature} \\
        \hline
        OpenAI & gpt-4o-mini & 0.4 \\
        Claude & claude-3-5-haiku-20241022 & 0.4 \\
        Claude & claude-3-7-sonnet-20250219 & 0.4 \\
        Gemini & gemini-1.5-pro & 0.4 \\
        Meta LLaMA & Llama-3.3-70B-Instruct-Turbo-Free & 0.4 \\
        DeepSeek & deepSeek-V3 & 0.4 \\
        Qwen & Qwen2.5-72B-Instruct-Turbo & 0.4 \\
        Grok & grok-2-1212 & 0.4 \\
        \hline
    \end{tabular}
\end{table}

\subsection{Custom Prompting Technique}
We developed a custom prompting strategy that leverages multiple techniques to optimize the performance and consistency of the LLMs in generating non-functional requirements (NFRs) from functional requirements. Our approach encompasses ten key techniques—detailed in Table~\ref{tab:prompting_techniques}—ranging from role assignment for domain expertise to tone and style direction for professional output.

This multi-faceted strategy ensures a consistent, thorough, and systematic approach to NFR generation, grounded in ISO/IEC 25010:2023 standards and tailored for maximum effectiveness. The complete custom prompt is available in our public repository \cite{HumanEvaluationDataset}.

\begin{table*}[h]
    \centering
    \caption{Prompting Techniques Used in the NFR Generation Prompt}
    \label{tab:prompting_techniques}
    \renewcommand{\arraystretch}{1.5}
    \begin{tabular}{p{3cm} p{7cm} p{6.5cm}}
        \toprule
        \textbf{Technique} & \textbf{Description} & \textbf{Purpose} \\
        \midrule
        Role Assignment & Assigns the LLM the role of a software quality engineer. & Ensures domain-specific expertise and technical precision in responses. \\
        Task Clarity and Specificity & Defines the task as generating NFRs aligned with ISO/IEC 25010 for given FRs. & Reduces ambiguity, ensuring the LLM understands the exact goal and scope. \\
        Structured Output Format & Specifies a JSON-like output structure. & Enforces consistency and usability, aligning with the user’s schema for processing. \\
        Example-Based Guidance & Provides a detailed example of NFRs for a sample FR. & Illustrates the expected format and reasoning, improving generalization to new inputs. \\
        Constraint Enforcement & Requires specific, testable NFRs with measurable thresholds, avoiding vagueness. & Ensures practical, developer-friendly requirements suitable for implementation and testing. \\
        Contextual Grounding & Anchors NFR generation in ISO/IEC 25010 categories and sub-characteristics. & Provides a standardized, comprehensive framework for quality attribute coverage. \\
        Iterative Refinement Instructions & Guides the LLM to analyze each FR and apply relevant attributes systematically. & Promotes efficiency and relevance by focusing only on applicable NFRs. \\
        Justification Requirement & Mandates a justification linking each NFR to the FR’s characteristics. & Enhances transparency and traceability, aiding validation by users. \\
        Input Flexibility & Allows handling a variable number of FRs individually. & Ensures scalability and adaptability to different user inputs. \\
        Tone and Style Direction & Uses technical language and examples to set a formal tone. & Aligns responses with software engineering needs, maintaining professionalism. \\
        \bottomrule
    \end{tabular}
\end{table*}

\subsection{Automated Pipeline}
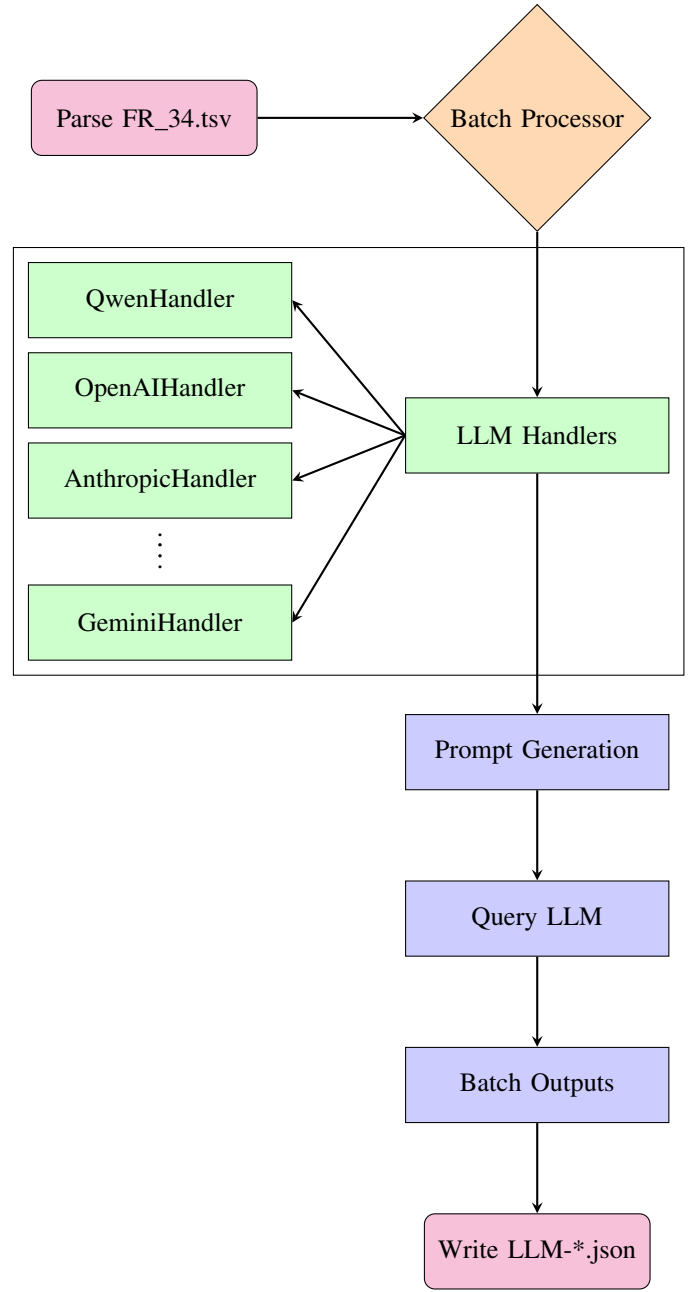
\begin{figure}[h]
    \centering
    \tikzstyle{startstop} = [rectangle, rounded corners, minimum width=3cm, minimum height=1cm, text centered, draw=black, fill=magenta!30]
\tikzstyle{process} = [rectangle, minimum width=3.5cm, minimum height=1cm, text centered, draw=black, fill=blue!20]
\tikzstyle{decision} = [diamond, minimum width=3cm, minimum height=1cm, text centered, draw=black, fill=orange!30]
\tikzstyle{subprocess} = [rectangle, minimum width=3.5cm, minimum height=1cm, text centered, draw=black, fill=green!20]
\tikzstyle{arrow} = [thick,->,>=stealth]

\begin{center}
\begin{tikzpicture}[node distance=1.2cm]

    \node (parse) [startstop] {Parse FR\_34.tsv};
    \node (batch) [decision, right=of parse, xshift=1cm] {Batch Processor};
    
    
    \begin{scope}[local bounding box=redbox]
        \node (llmhandlers) [subprocess, below=of batch, yshift=-1cm] {LLM Handlers};
        
        \node (qwen) [subprocess, left=of llmhandlers, xshift=-0.3cm, yshift=1.8cm] {QwenHandler};
        \node (openai) [subprocess, left=of llmhandlers, xshift=-0.3cm, yshift=0.6cm] {OpenAIHandler};
        \node (anthropic) [subprocess, left=of llmhandlers, xshift=-0.3cm, yshift=-0.6cm] {AnthropicHandler};
        \node (gemini) [subprocess, left=of llmhandlers, xshift=-0.3cm, yshift=-2.49cm] {GeminiHandler};

        \node at ($(anthropic)!.4!(gemini)$) {\vdots};
        \node at ($(anthropic)!.474!(gemini)$) {\vdots};
    \end{scope}
    \draw[black] ($(redbox.south west) - (2mm,2mm)$)
        rectangle ($(redbox.north east) + (2mm,2mm)$);
        
    \node (prompt) [process, below=of llmhandlers, yshift=-2cm] {Prompt Generation};
    \node (query) [process, below=of prompt] {Query LLM};
    \node (batchout) [process, below=of query] {Batch Outputs};
    \node (write) [startstop, below=of batchout] {Write LLM-*.json};

    \draw [arrow] (parse.east) -- (batch.west);
    
    \draw [arrow] (batch.south) -- (llmhandlers.north);
    \draw [arrow] (llmhandlers.west) -- (qwen.east);
    \draw [arrow] (llmhandlers.west) -- (openai.east);
    \draw [arrow] (llmhandlers.west) -- (anthropic.east);
    \draw [arrow] (llmhandlers.west) -- (gemini.east);

    \draw [arrow] (llmhandlers) -- (prompt.north);

    \draw [arrow] (prompt) -- (query);
    \draw [arrow] (query) -- (batchout);
    \draw [arrow] (batchout) -- (write);

\end{tikzpicture}
\end{center}
    \caption{Deno-based Pipeline Architecture for Automated LLM-driven NFR Generation}
    \label{fig:deno_pipeline_arch}
\end{figure}

A Deno-based pipeline (illustrated in Figure \ref{fig:deno_pipeline_arch}) has been developed to automate the process of calling LLMs through APIs, collecting their responses, and storing the extracted NFRs for subsequent analysis. 

This automated pipeline ensures consistency and reproducibility in multiple experimental runs. In addition, the system also includes a sampling mechanism that extracts representative outputs for validation by human evaluators.
\section{Evaluation Framework} \label{sec:evaluation}

We assess the quality of Non-Functional Requirements (NFRs) generated from the selected Functional Requirements (FRs) through a dual-evaluation strategy that relies on human expertise. This approach comprises two distinct tasks: an ordinal classification to score the generated NFRs and an attribute selection process to validate their assigned quality attributes. To support these evaluations, we built a web-based application using SvelteKit, enabling 10 evaluators—experienced professionals averaging 13 years in roles such as Senior Software Engineer, Principal Program Manager, and Software Quality Engineer (ranging from 2 to 30 years of experience)—to submit their assessments. This section details the methodologies, criteria for scoring and attribute selection, evaluator assignments, and the use of the ISO/IEC 25010:2023 standard quality model.

\subsection{Evaluation Methodology for NFR Scoring}
\label{subsec:nfr_scoring}
For each of the 34 FRs selected from the dataset (see Section~\ref{sec:data_collection}), we tasked the LLMs described in Section~\ref{sec:llm_selection} with generating applicable NFRs. Since a single FR can be associated with multiple non-functional concerns, this ensures a more comprehensive assessment of relevant quality attributes. Each generated NFR is linked to a quality attribute from the ISO/IEC 25010:2023 standard \cite{ISO25010}, and human evaluators assess their applicability and validity. The evaluation involves assigning two ordinal scores, ranging from 1 to 5, to each NFR:
\begin{enumerate}
    \item \textbf{Applicability}: Evaluates the relevance and suitability of the ISO/IEC 25010:2023 quality attribute assigned to the NFR in the context of its corresponding FR.
    \item \textbf{Validity}: Assesses the coherence, relevance, and justification of the NFR with respect to its corresponding FR.
\end{enumerate}

\begin{figure}[ht]
    \centering
    \includegraphics[width=1\linewidth]{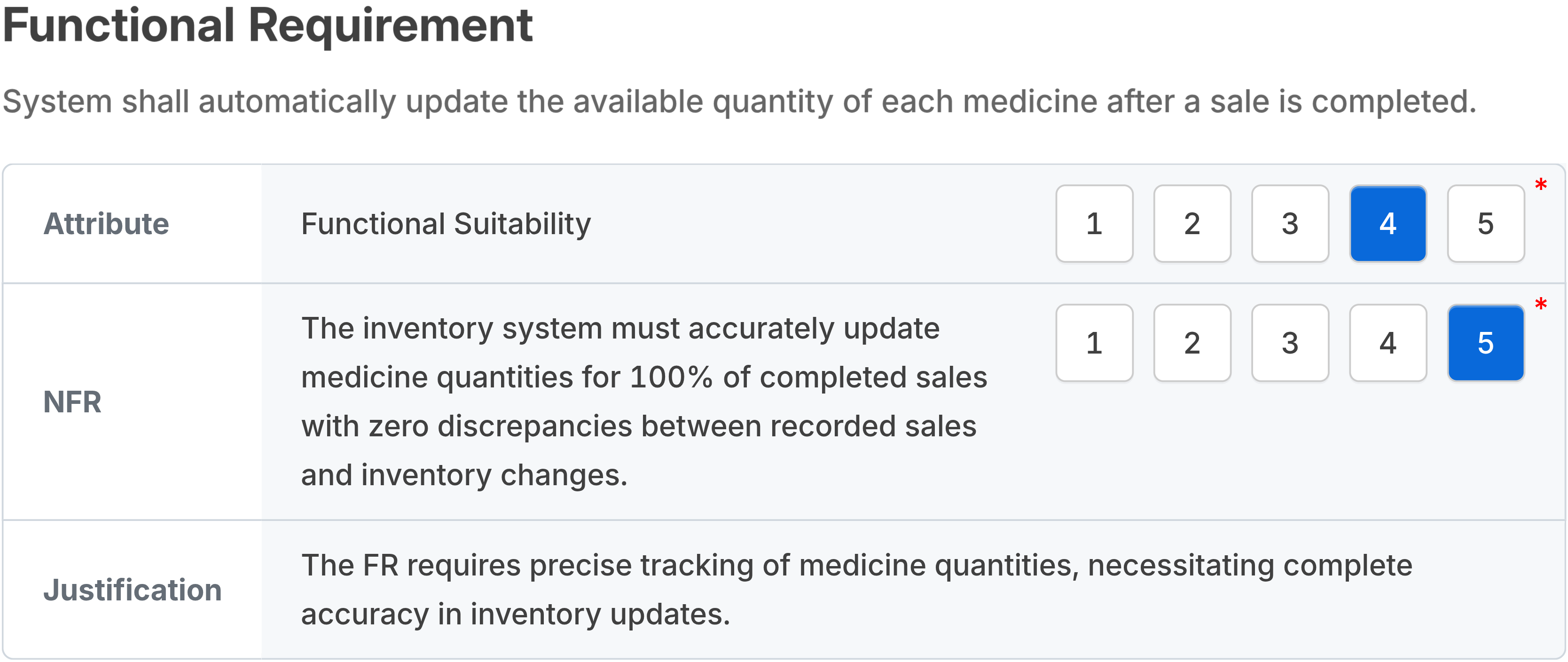}
    \caption{Human Evaluation Interface for NFR Validity and Applicability Scoring}
    \label{fig:nfr_evaluation}
\end{figure}

Each of the 10 evaluators was pre-assigned three FRs from the set of 34, along with all NFRs generated by their designated LLM. The evaluators accessed the web interface, as shown in Figure~\ref{fig:nfr_evaluation}, to review each FR, its generated NFRs, and their associated quality attributes, providing scores based on the criteria outlined below (subsection \ref{subsec:criteria}).

\subsection{Scoring Criteria} \label{subsec:criteria}
\subsubsection{Applicability of Quality Attributes}
Each NFR is associated with a quality attribute defined in the ISO / IEC 25010:2023 standard quality model \cite{ISO25010}, which includes Functional Suitability, Performance Efficiency, Compatibility, Usability, Reliability, Security, Maintainability, Flexibility, and Safety. The applicability score evaluates how relevant and suitable the assigned attribute is for the corresponding FR, ensuring a meaningful inference process and proper alignment with the generated NFR. The scoring rubric is as follows:
\begin{itemize}
    \item \textbf{1 - Not Applicable}: Completely unsuitable and unrelated to the NFR or FR.
    \item \textbf{2 - Barely Applicable}: Minimal or tangential relevance to the FR.
    \item \textbf{3 - Somewhat Applicable}: Moderately relevant, but with questionable fit.
    \item \textbf{4 - Mostly Applicable}: Strong fit with minor doubts or edge cases.
    \item \textbf{5 - Perfectly Applicable}: Exact and ideal match for the NFR and FR.
\end{itemize}

\subsubsection{Validity of NFRs}
The validity score reflects the quality and appropriateness of each NFR generated for its corresponding FR. Evaluators assign a score from 1 to 5 based on the following rubric:
\begin{itemize}
    \item \textbf{1 - Invalid}: Incoherent, irrelevant, or contradictory to the FR.
    \item \textbf{2 - Barely Valid}: Major flaws present, only partially relevant to the FR.
    \item \textbf{3 - Partially Valid}: Somewhat clear but with noticeable issues or inconsistencies.
    \item \textbf{4 - Mostly Valid}: Clear and relevant, with only minor flaws.
    \item \textbf{5 - Fully Valid}: Specific, achievable, and perfectly justified in the context of the FR.
\end{itemize}

\subsection{Evaluation Methodology for Attribute Selection}
\label{subsec:attribute_selection}
In addition to scoring NFRs, we conducted a second evaluation to validate the quality attributes assigned to the generated NFRs. This evaluation aimed to determine whether LLMs generated NFRs that correctly belonged to the intended classification category, rather than hallucinating responses, such as producing misclassified or unrelated NFRs (for example, assigning a security-related NFR under usability). By comparing the human-selected attributes with those assigned by the LLMs, we evaluated the models' ability to generate NFRs that align with the expected quality categories.

\begin{figure}[ht]
    \centering
    \includegraphics[width=1\linewidth]{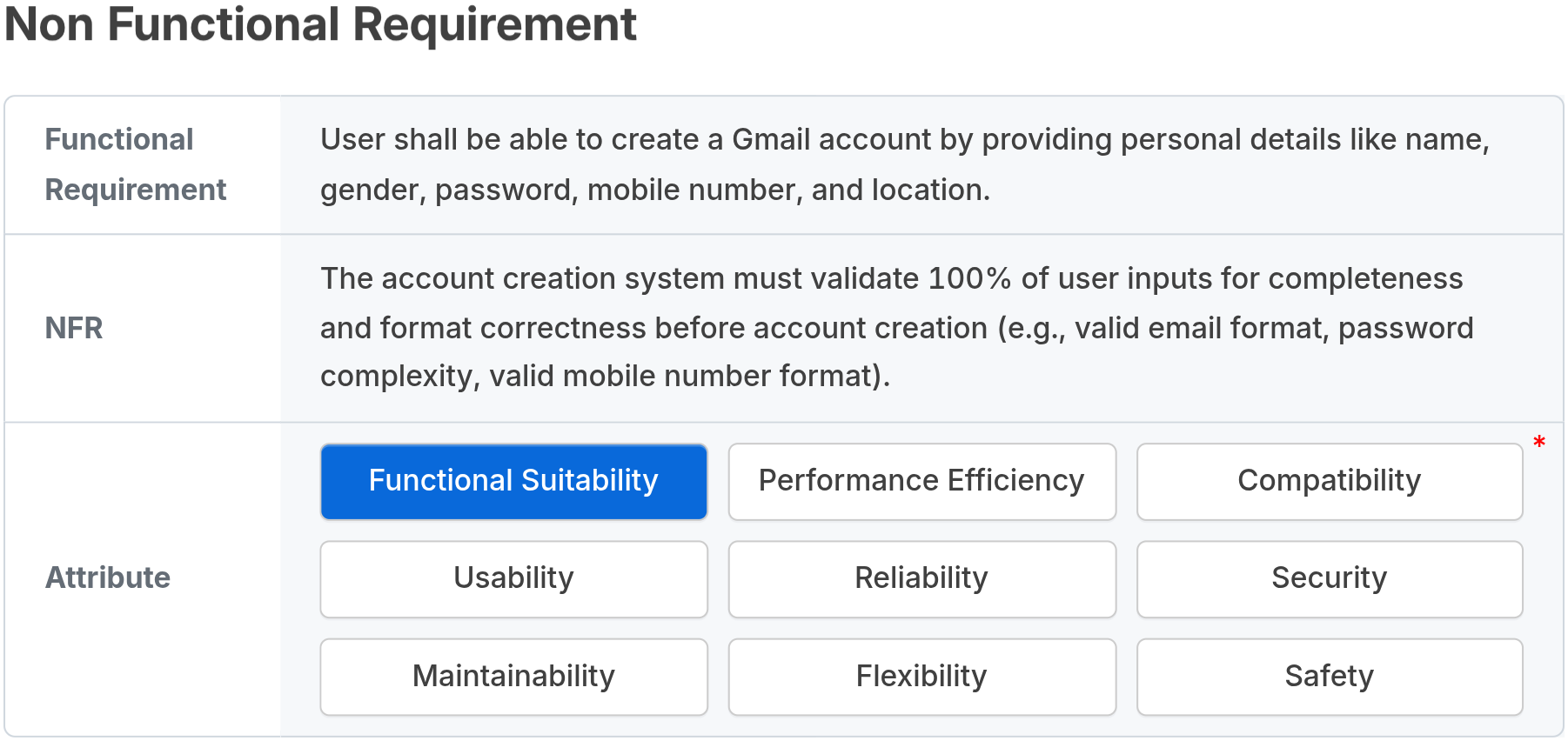}
    \caption{Human Evaluation Interface for ISO/IEC 25010 Attribute Selection.}
    \label{fig:attr_evaluation}
\end{figure}

Evaluators were presented with a different set of FR-NFR pairs, as shown in Figure~\ref{fig:attr_evaluation}, and asked to select the most appropriate quality attribute from the nine options of ISO/IEC 25010:2023. Importantly, this was a blind review: evaluators did not have prior access to the attributes assigned by the LLMs, ensuring an unbiased assessment. 

To ensure that there is no overlap between the two evaluations, the FRs used in this attribute selection task are disjoint from those used in the NFR scoring task. Each evaluator was pre-assigned three FRs, and for each FR, they reviewed all generated NFRs along with their LLM-assigned attributes. The evaluators selected one attribute per NFR from the provided list, without assigning a numerical score.

\subsection{Data Collection Process}
The web application supported both evaluation tasks. For the NFR scoring task, the assessors reviewed their pre-assigned FRs, all generated NFRs, and associated ISO/IEC 25010:2023 attributes, submitting scores (1–5) for validity and applicability through the web interface. The evaluators reviewed a different set of FR-NFR pairs for the attribute selection task and selected one attribute per NFR from the nine options. All responses were recorded in a centralized SQLite database for subsequent analysis, and the dataset has been made openly available at \cite{HumanEvaluationDataset}. To minimize bias, the evaluators were instructed to base their judgments solely on the provided criteria, without knowing the specific LLM that generates each NFR.

\subsection{Expected Outcomes}
The dual-evaluation framework enables a comprehensive assessment of the quality of NFR in multiple dimensions. The NFR scoring task quantifies the validity and applicability of the generated NFRs, allowing for the potential variability in the number of NFRs per FR. This will provide insights into the LLMs' ability to identify high-quality non-functional aspects. The attribute selection task offers a complementary perspective by validating the LLM-assigned quality attributes against human preferences, highlighting potential discrepancies in attribute assignment.

\section{Results} \label{sec:results}
A total of 1,593 NFRs were generated for the 34 FRs across the eight LLMs. However, due to the time-intensive nature of expert evaluation, we limited our study to a manageable subset for in-depth analysis. Specifically, two main tasks were conducted: \emph{NFR Scoring} (Validity and Applicability) and \emph{Attribute Selection}. Accordingly, 174 NFRs were selected for scoring, and 168 for attribute accuracy review, using a stratified sampling approach to ensure representation across all eight LLMs. The following findings emerged from these evaluations.

\subsection{NFR Scoring}

\subsubsection{Validity}
\begin{figure}[ht]
    \centering
    \includegraphics[width=1\linewidth]{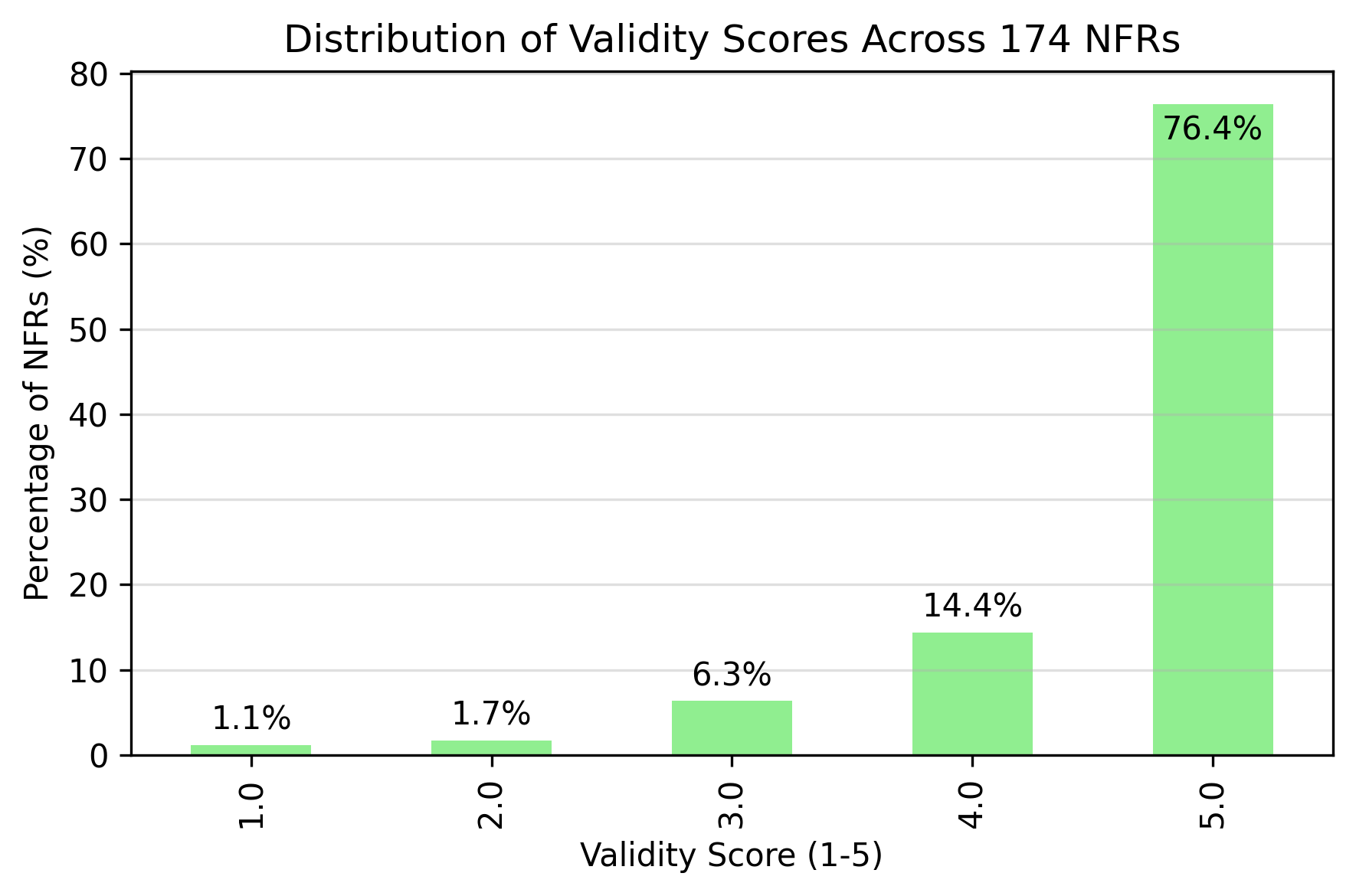}
    \caption{Distribution of NFR Validity Scores.}
    \label{fig:validity_distribution}
\end{figure}

For the 174 selected NFRs, evaluators assigned a median validity score of 5.0 (mean: 4.63). Figure \ref{fig:validity_distribution} illustrates the score distribution.
\begin{itemize}
    \item \textbf{High Validity (4-5):} 90.8\% of NFRs received a validity score of 4 or higher, with 14.4\% rated at 4 and 76.4\% rated at 5. This indicates that the vast majority of generated NFRs were coherent and closely aligned with their corresponding FRs, demonstrating strong relevance and clarity.
    \item \textbf{Moderate Validity (3):} 6.3\% required minor clarifications but were generally relevant.
    \item \textbf{Low Validity (1-2):} 2.9\% either lacked clear relevance or contradicted the FR, reflecting rare but notable failures.
\end{itemize}

\subsubsection{Applicability}  
\begin{figure}[ht]
    \centering
    \includegraphics[width=1\linewidth]{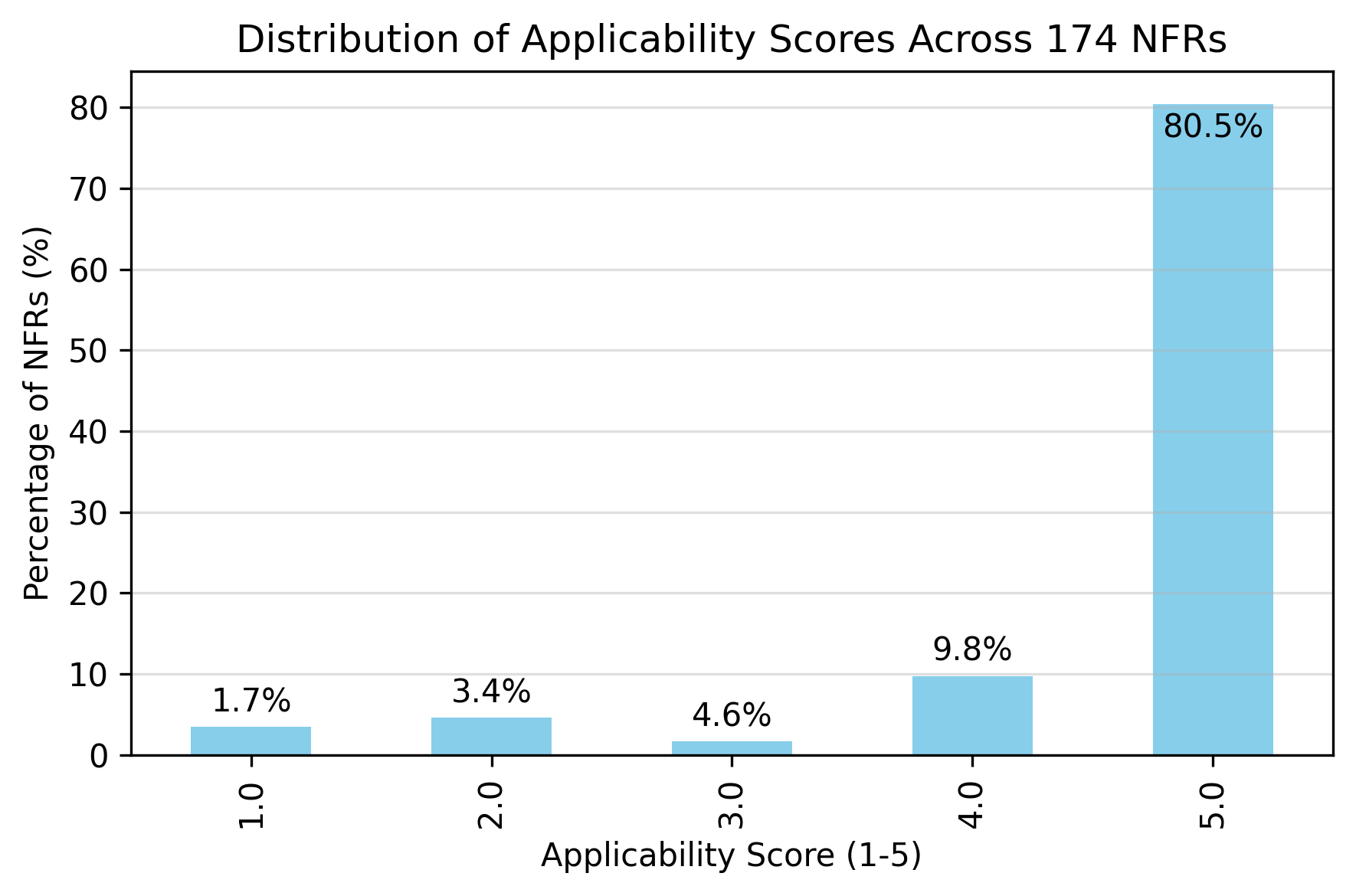}
    \caption{Distribution of NFR Applicability Scores.}
    \label{fig:applicability_distribution}
\end{figure}

The Applicability scores assess how relevant and suitable the assigned ISO/IEC 25010:2023 attribute is for the corresponding FR, ensuring meaningful alignment with the generated NFR. The median score was 5.0 (mean: 4.59), indicating a strong trend towards selecting relevant attributes for NFR generation. Figure~\ref{fig:applicability_distribution} illustrates the distribution of these scores, highlighting the proportion of NFRs that were classified as highly, moderately, or barely applicable.

\begin{itemize}  
    \item \textbf{Highly Applicable (4-5):}  90.2\% were strong or exact attribute matches, suggesting LLMs frequently assign relevant categories.
    \item \textbf{Moderately Applicable (3):} 1.7\% displayed partial alignment but required refinement.
   \item \textbf{Barely/Not Applicable (1-2):} 8.0\% of NFRs were assigned attributes with minimal or no relevance to their corresponding FRs, suggesting misclassification, often due to overly broad or tangential attributes. 
\end{itemize}

\subsection{Attribute Selection}
\begin{figure}[hbt!]
    \centering
    \includegraphics[width=1\linewidth]{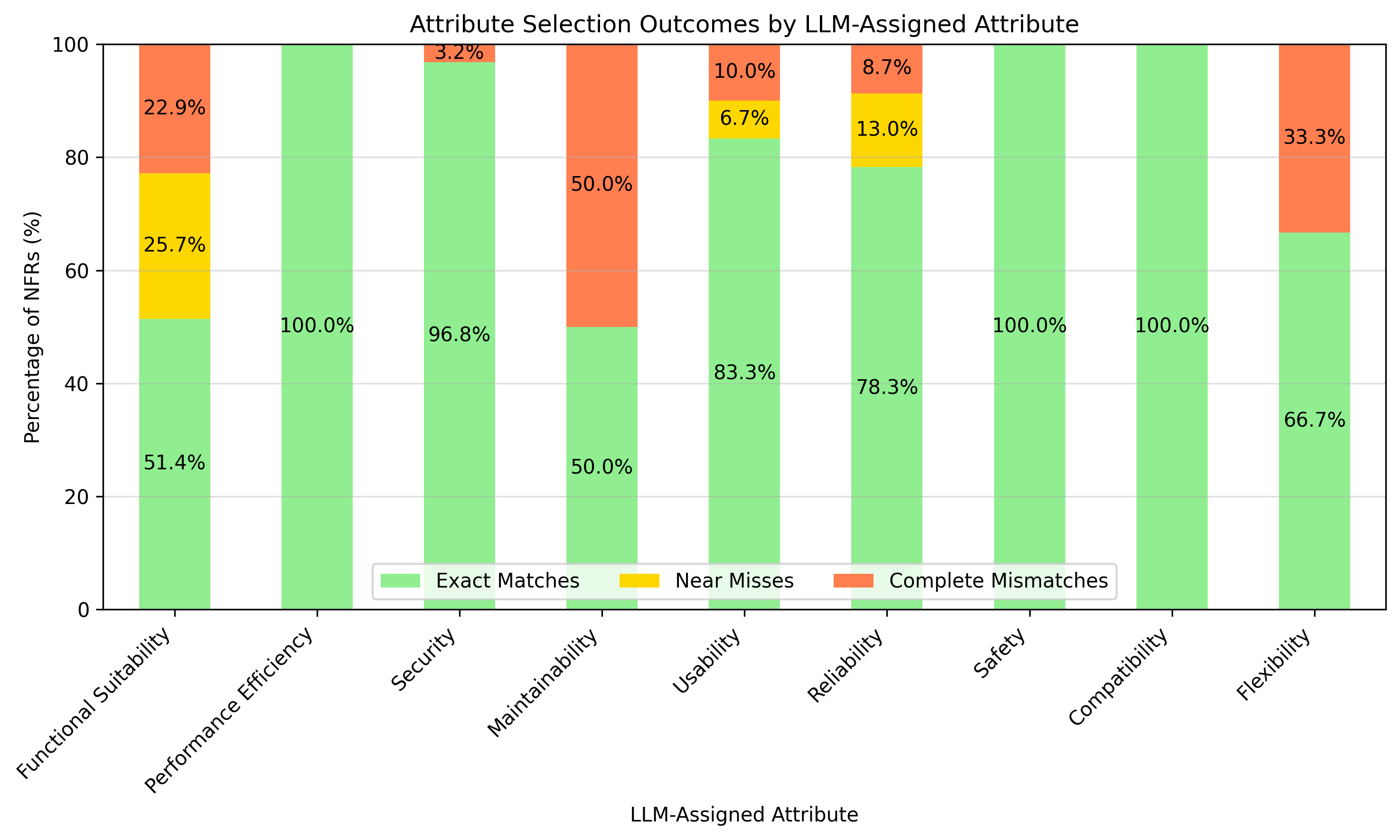}
    \caption{Comparison of Expert-Selected vs. LLM-Assigned Attributes.}
    \label{fig:attribute_selection-stacked}
\end{figure}

In the separate attribute selection task, experts re-evaluated 168 pairs of NFR attributes to assess the alignment between LLM-assigned and expert-selected categories. As illustrated in Figure~\ref{fig:attribute_selection-stacked}:
\begin{itemize}
    \item \textbf{Exact Matches:} 80.4\% (135/168) of the attributes the LLMs chose aligned exactly with expert selections, a strong indicator of categorization reliability.
    \item \textbf{Near Misses:} 8.3\% (14/168) involved related but different attributes (e.g. \emph{Performance Efficiency} vs. \emph{Reliability}), often differing in nuance rather than intent.
    \item \textbf{Complete Mismatch:} 11.3\% (19/168) demonstrated a complete mismatch, where the assigned attributes were inconsistent or irrelevant, such as \emph{Functional Suitability} mislabeled as \emph{Security}.
\end{itemize}

\begin{figure}[hbt!]
    \centering
    \includegraphics[width=0.9\linewidth]{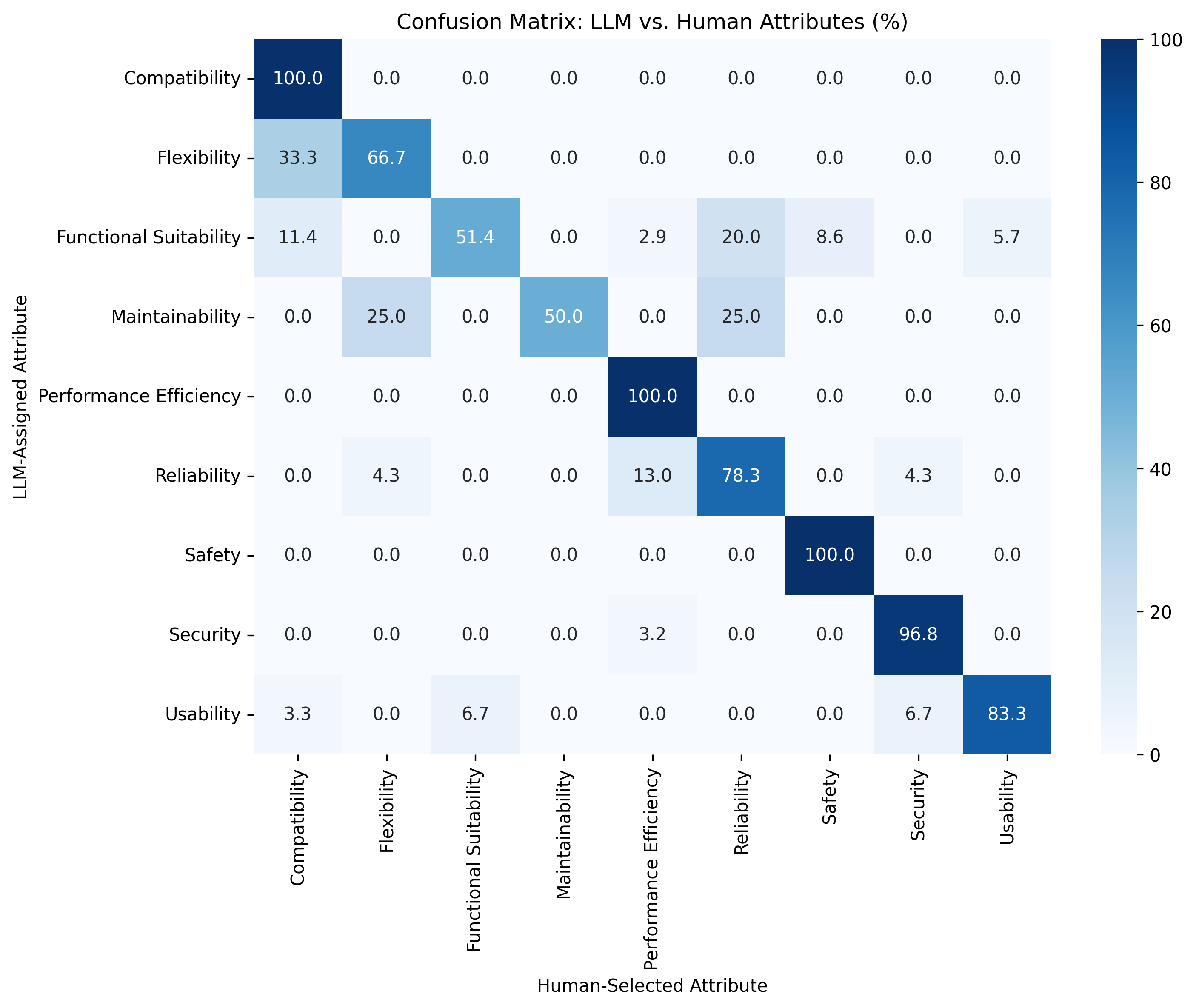}
    \caption{Confusion Matrix: LLM-Assigned vs. Expert-Selected Attributes.}
    \label{fig:attribute_selection-matrix}
\end{figure}

Figure~\ref{fig:attribute_selection-matrix} provides further insight into the mismatched cases between LLM-assigned and human-selected attributes. While Performance Efficiency and Compatibility were always correctly classified, certain misclassification patterns emerged. For instance, 20\% of NFRs generated under Functional Suitability by the LLMs were actually categorized as Reliability by human evaluators, indicating that the models sometimes confuse functional effectiveness with system dependability. Similarly, 33.3\% of NFRs labeled as Flexibility by the LLM were actually Compatibility, suggesting difficulty in distinguishing between system adaptability and interoperability. These findings highlight areas where LLMs may struggle to infer the appropriate quality attribute when generating NFRs, underscoring the need to refine LLM-based NFR classification for better alignment with human judgment.


Some of these mismatches, along with variations in Validity and Applicability scores, may be more pronounced in specific LLMs. To further analyze these differences, Table~\ref{tab:llm_comparison} presents a comparative overview of the results across the eight evaluated LLMs, highlighting trends in NFR generation accuracy and classification performance.

\begin{table}[hbt!]
    \centering
    \caption{Validity, Applicability, and Attribute Accuracy Per LLM Performance}
    \label{tab:llm_comparison}
    \begin{tabular}{l p{1.3cm} l p{1.3cm}}
        \toprule
        \textbf{LLM} & \textbf{Avg. Validity} & \textbf{Avg. Applicability} & \textbf{Attr. Accuracy (\%)} \\
        \midrule
        gpt-4o-mini & 4.55 & 4.91 & 75.0 \\
        claude-3-5-haiku & 4.66 & 4.74 & 85.3 \\
        claude-3-7-sonnet & 3.96 & 3.67 & 74.1 \\
        gemini-1.5-pro & 4.79 & 4.64 & 90.9 \\
        llama-3.3-70B & 4.94 & 4.94 & 84.2 \\
        deepSeek-V3 & 4.69 & 4.88 & 71.4 \\
        Qwen2.5-72B & 4.72 & 4.20 & 80.0 \\
        grok-2 & 4.81 & 4.97 & 82.6 \\
        \bottomrule
    \end{tabular}
\end{table}

\section{Discussion} \label{sec:discussion}
Our results, based on expert evaluations of 174 NFRs for validity and applicability and 168 NFRs for attribute accuracy, suggest that LLMs can effectively generate non-functional requirements from functional requirements. The evaluations, measured on a scale of 1.0 to 5.0, achieved a mean validity score of 4.63, a mean applicability score of 4.59, and an overall attribute accuracy of 80.4\%. These metrics indicate that LLMs produce NFRs that are largely coherent, relevant, and appropriately categorized, supporting their practical utility in requirements elicitation. Table~\ref{tab:llm_comparison} summarizes LLM performance, showing consistent quality across models, though attribute classification errors in 19.6\% of cases (near misses and mismatches) highlight areas for refinement.

\subsection{Research Questions Addressed}

\textbf{RQ1: How effective are LLMs in generating NFRs that are clear, relevant, and comprehensive?}\\
The mean validity of 4.63 and applicability of 4.59, with 90.8\% and 90.2\% of NFRs scoring $\geq$ 4 (Fig.~\ref{fig:validity_distribution} and Fig.~\ref{fig:applicability_distribution}), demonstrate high effectiveness. Most NFRs are clear and relevant, as seen in high-scoring examples (e.g., `grok-2`: 5.0/5.0 for \emph{Functional Suitability}). However, 2.9\% of NFRs scored low in validity (1-2), such as `Qwen2.5-72B` (1.0/1.0), indicating occasional gaps in comprehensiveness that may require human review. While direct comprehensiveness scoring was not conducted, our approach aimed to elicit all \emph{applicable} NFRs based on ISO/IEC 25010 categories, implying an attempt at broad coverage within the defined quality framework.

\textbf{RQ2: Among the eight LLMs tested, which model delivers the most reliable NFR generation according to the ISO/IEC 25010:2023 criteria?}\\
Table~\ref{tab:llm_comparison} provides a comparative overview of LLM performance. 
No statistically significant difference in attribute accuracy was measured across models; however, \texttt{gemini-1.5-pro} achieved the highest attribute accuracy, and \texttt{llama-3.3-70B} showed marginally superior validity and applicability scores. The limited sample size restricts our ability to definitively identify a single top-performing model. Future work with larger datasets is necessary to confirm these trends and establish statistically robust model comparisons.

\textbf{RQ3: Can a custom prompting technique, enhanced with role descriptions and in-context learning, improve the precision of NFR generation compared to baseline methods?}
Initial experiments with a basic prompt that used just Role Assignment and Contextual Grounding techniques yielded inconsistent outputs, lacking specificity and measurable criteria. To address this, we developed an advanced prompt incorporating techniques such as Constraint Enforcement (requires testable NFRs) and example-based guidance (Table~\ref{tab:prompting_techniques}). While a formal baseline comparison was not conducted, the advanced prompt produced NFRs with 80.4\% attribute accuracy and 90\%+ high-quality scores (4-5), suggesting improved precision over the initial prompt’s anecdotal inconsistency. For instance, `grok-2' consistently generated testable NFRs (e.g., 5.0/5.0), unlike earlier vague outputs. In future work, we plan to systematically evaluate various prompting variants to quantify their impact on generation precision.

\subsection{Implications}
The 90.8\% validity and 80.4\% attribute accuracy suggest LLMs can automate NFR elicitation in over 80\% of cases, reducing manual effort significantly. However, 11.3\% complete mismatches (e.g., \emph{Reliability} as \emph{Performance Efficiency}, Fig.~\ref{fig:attribute_selection-stacked}) and 8.3\% near misses (e.g., \emph{Usability} vs. \emph{Functional Suitability}) indicate that abstract attributes like \emph{Functional Suitability} (51.4\% accuracy) require refinement.

These findings point to two key considerations: (1) advanced prompting techniques and richer in-context examples, as shown in RQ3, enhance precision and could further reduce misalignment; (2) human oversight remains essential for the 19.6\% of cases with attribute errors, particularly in safety-critical domains. LLMs generally show promise for NFR elicitation if supported with human oversight and iterative feedback.

\section{Limitations}
Although our study provides valuable information on the use of LLMs for automated generation of NFRs, it is important to recognize its limitations.  First, the sample size of functional requirements (34 FRs) and NFR evaluations (174 for scoring, 168 for attribute selection) is relatively modest. Although this size is representative of FR counts in midsize systems, as derived from our analysis of SRS documents, a larger and more diverse dataset could provide more robust and statistically significant results, particularly for discerning performance differences between LLMs.

Secondly, the dataset characteristics may influence the generalizability of our findings. The FR\_NFR\_dataset, while valuable, is derived from SRS documents within a specific time frame (2008-2011) and may not fully represent the current diversity and complexity of software requirements in contemporary software development.  Furthermore, the data set itself, while manually curated, can contain inherent biases or inconsistencies in the NFR classification, which could affect the baseline evaluation.

Third, although we employed experienced professionals as evaluators to mitigate subjectivity, the inherent biases of human judgment cannot be completely eliminated.  The evaluators' backgrounds, while relevant, may introduce specific perspectives in assessing NFR validity and attribute applicability.  Future studies could benefit from involving a larger and more diverse group of evaluators, and potentially incorporating inter-rater reliability measures to further quantify and address potential evaluator bias.

Finally, our study focused on generating NFRs from FRs in isolation.  In practice, NFR elicitation is a more complex process that involves stakeholder interaction, consideration of business goals, and iterative refinement.  Our methodology, while providing a focused evaluation of LLM capabilities in this specific generation task, does not fully replicate the broader context of real-world NFR elicitation.

\section{Conclusion}  \label{sec:conclusion}
This work investigated the use of LLMs for automating the generation of non-functional requirements (NFRs) from functional requirements. Our contributions are multifaceted. We developed a system for NFR generation, leveraging a prompting pipeline aligned with the ISO/IEC 25010:2023 standard. In total, 1,593 NFRs were generated from 34 functional requirements using eight different LLMs.

For a subset of 174 NFRs, domain experts assigned validity and applicability scores ranging from 1 to 5, with median scores of 5.0 (means: 4.63 for validity and 4.59 for applicability), indicating a trend toward relevant, high-quality NFRs and essential coverage. In a separate attribute selection task (168 NFRs), 80.4\% of LLM-assigned attributes exactly matched expert decisions, with 8.3\% near misses and 11.3\% complete mismatches, underscoring the system's reliability. A comparative evaluation of the eight LLMs revealed variations in performance. \texttt{gemini-1.5-pro} exhibited the highest attribute accuracy, while \texttt{llama-3.3-70B} achieved slightly higher validity and applicability scores. These variations suggest that certain models may be better suited for different aspects of NFR generation. 

However, the modest sample size, potential human evaluator subjectivity, and the isolation of NFR generation from stakeholder interaction mean the study does not fully represent real-world elicitation processes. Overall, this work contributes to the broader discussion of AI-assisted requirements engineering by demonstrating how LLMs can support and enhance NFR generation. Expanding the dataset and conducting a case study in a real setting will be a key next step to enable a more robust and statistically meaningful assessment of the generated requirements.

\section{Future Work: Research Agenda} \label{sec:future}

Considering that this is a research topic with significant unexplored potential, we outline our future work as a research agenda, highlighting multiple areas for further investigation and improvement. 


\subsection{Increasing Sample Size for Validation}
Although our horizontal evaluation provided valuable insights, the limited sample size—constrained by expert availability—reduces the statistical power of our findings. As a next step, we plan to transition from a broad horizontal evaluation to a more in-depth vertical analysis by increasing the validation dataset. This expansion will enable a more robust and comprehensive assessment of the generated requirements.

\subsection{Benchmark Development}
While we have already provided a dataset of annotated FRs and NFRs \cite{HumanEvaluationDataset}, future work will focus on expanding it with additional expert evaluations to enhance its reliability and usability as a benchmark. This enriched dataset will not only support further research in automating requirements elicitation but also provide a standardized reference for comparative studies on NFR generation. Additionally, we aim to explore advanced prompting techniques and integrate vulnerability detection approaches to further refine the quality and applicability of generated requirements.

\subsection{Generating Non-functional Requirements for AI-enabled Systems}  
AI/ML-enabled systems require non-functional requirements for unique quality attributes such as transparency, ethics, and fairness \cite{villamizar2021requirements} \cite{de2025classification}. Defining and measuring these NFRs pose challenges, necessitating structured approaches for their effective specification \cite{habibullah2024scoping} \cite{damirchi2023non}. In future work, we plan to explore how our approach can support the generation and specification of NFRs tailored to these needs, aiding AI/ML systems engineers in aligning elicitation with critical requirements.

\subsection{Evaluation of Individual and Group Requirements}  

As described in ISO/IEC/IEEE 29148 \cite{8559686}, the evaluation of generated non-functional requirements (NFRs) needs also consider their role within a complete set of requirements. Although individual NFRs must adhere to well-defined quality attributes such as feasibility, completeness, and verifiability, requirements in a group must \cite{8559686}:  

\begin{itemize}  
    \item Be complete: The set of requirements must adequately describe all necessary characteristics without leaving critical gaps.  
    \item Be consistent: No two requirements should contradict or overlap in a way that creates ambiguity.  
    \item Be feasible as a set: The complete set should be achievable under given system constraints, including cost and technical limitations.  
    \item Be able to be validated: The set of requirements should provide a structured and testable framework that aligns with the system's goals.  
\end{itemize}  

Following existing validation approaches \cite{porter2025requirements}, our objective is to explore automated techniques to evaluate individual and grouped requirements, extending beyond human-likeness assessments. This includes leveraging formalized validation strategies to check whether individual NFRs comply with best practices in requirements engineering while ensuring that sets of NFRs do not introduce inconsistencies or gaps in coverage.

\subsection{Improvement of Requirements Attributes}  

In addition to type classification, which is essential for distinguishing different types of NFRs to ensure proper allocation and prioritization, our approach also aims to improve the practical applicability of generated NFRs by incorporating additional attributes \cite{8559686}, including:

\begin{itemize}
\item Rationale: Capturing the reasoning behind each requirement, referencing supporting analysis, trade-offs, or simulations to justify its necessity.
\item Difficulty level: Categorizing requirements based on their complexity (e.g., easy, moderate, difficult) to assist in cost estimation and project planning.
\end{itemize}

\subsection{Human-in-the-Loop Evaluation and Task Allocation}  

Beyond automated validation and structural assessment, it is crucial to evaluate how requirements engineers perceive generated NFRs and how they interact with AI-assisted tools. Our initial validation compared expert-written and LLM-generated NFRs to avoid bias. However, more research is needed to understand the usefulness, clarity, and relevance of AI-generated NFRs in real-world requirements engineering.  

To address this, we propose a human-in-the-loop approach, where expert feedback and iterative refinements enhance the quality and usability of automatically generated requirements. This interaction will ensure that AI-generated NFRs align with industry expectations and can effectively support the elicitation process. Future work will explore qualitative and quantitative feedback from practitioners, assessing:  

\begin{itemize}  
    \item Whether the generated NFRs align with real-world expectations.  
    \item How effectively do they support the elicitation process  
    \item The perceived effort required to refine or adapt them for actual projects.  
\end{itemize}  

Additionally, our aim is to investigate the selection of LLMs for task allocation in the requirements elicitation process. As proposed in \cite{nascimento2021approach}, AI-driven task allocation can optimize workflow efficiency by selecting the most suitable LLM based on developer expertise and task requirements. Incorporating varying levels of automation \cite{melo2023identifying} will help balance AI-generated content with human oversight, ensuring that the tool adapts to different levels of user experience and project needs.  

\subsection{Framework Flexibility: Expanding Standard-Guided Inference}
Our framework integrates ISO/IEC 25010 as a guiding standard for generating non-functional requirements (NFRs), systematically aligning functional requirements with relevant quality attributes. This modular approach ensures structured, repeatable, and industry-aligned NFR generation.

However, ISO/IEC 25010 is not the only applicable standard for NFR elicitation and specification. In future work, the framework will be extended to support additional standards, including ISO/IEC/IEEE 29148, ISO/IEC 27001:2022 \cite{iso2022iso} for safety-critical applications, and ISO/IEC TS 25058:2024 \cite{ISO25058} for AI systems  engineering, enhancing its flexibility and applicability across diverse software domains.


\bibliographystyle{IEEEtran}
\bibliography{references}

\begin{thebibliography}{10}
\providecommand{\url}[1]{#1}
\csname url@samestyle\endcsname
\providecommand{\newblock}{\relax}
\providecommand{\bibinfo}[2]{#2}
\providecommand{\BIBentrySTDinterwordspacing}{\spaceskip=0pt\relax}
\providecommand{\BIBentryALTinterwordstretchfactor}{4}
\providecommand{\BIBentryALTinterwordspacing}{\spaceskip=\fontdimen2\font plus
\BIBentryALTinterwordstretchfactor\fontdimen3\font minus \fontdimen4\font\relax}
\providecommand{\BIBforeignlanguage}[2]{{%
\expandafter\ifx\csname l@#1\endcsname\relax
\typeout{** WARNING: IEEEtran.bst: No hyphenation pattern has been}%
\typeout{** loaded for the language `#1'. Using the pattern for}%
\typeout{** the default language instead.}%
\else
\language=\csname l@#1\endcsname
\fi
#2}}
\providecommand{\BIBdecl}{\relax}
\BIBdecl

\bibitem{mizouni2010towards}
R.~Mizouni and A.~Salah, ``Towards a framework for estimating system nfrs on behavioral models,'' \emph{Knowledge-Based Systems}, vol.~23, no.~7, pp. 721--731, 2010.

\bibitem{LaplanteKassab}
P.~A. Laplante and M.~Kassab, \emph{Requirements engineering for software and systems}.\hskip 1em plus 0.5em minus 0.4em\relax Auerbach Publications, 2022.

\bibitem{li2015stakeholder}
F.-L. Li, J.~Horkoff, A.~Borgida, G.~Guizzardi, L.~Liu, and J.~Mylopoulos, ``From stakeholder requirements to formal specifications through refinement,'' in \emph{Requirements Engineering: Foundation for Software Quality: 21st International Working Conference, REFSQ 2015, Essen, Germany, March 23-26, 2015. Proceedings 21}.\hskip 1em plus 0.5em minus 0.4em\relax Springer, 2015, pp. 164--180.

\bibitem{Brown2020}
T.~Brown, B.~Mann, N.~Ryder, M.~Subbiah, J.~D. Kaplan, P.~Dhariwal, A.~Neelakantan, P.~Shyam, G.~Sastry, A.~Askell \emph{et~al.}, ``Language models are few-shot learners,'' \emph{Advances in neural information processing systems}, vol.~33, pp. 1877--1901, 2020.

\bibitem{cheng2024generative}
H.~Cheng, J.~H. Husen, Y.~Lu, T.~Racharak, N.~Yoshioka, N.~Ubayashi, and H.~Washizaki, ``Generative ai for requirements engineering: A systematic literature review,'' \emph{arXiv preprint arXiv:2409.06741}, 2024.

\bibitem{porter2025requirements}
D.~Porter, J.~F. DeFranco, and P.~Laplante, ``Requirements specification automated quality analysis: Past, present, and future,'' \emph{Computer}, vol.~58, no.~1, pp. 101--104, 2025.

\bibitem{lano2024introduction}
K.~Lano, S.~Rahimi, S.~Tehrani, L.~Burgue{\~n}o, and M.~A. Umar, ``Introduction to theme section on requirements formalisation,'' \emph{Software and Systems Modeling}, vol.~23, no.~6, pp. 1451--1453, 2024.

\bibitem{8559686}
``Iso/iec/ieee international standard - systems and software engineering -- life cycle processes -- requirements engineering,'' \emph{ISO/IEC/IEEE 29148:2018(E)}, pp. 1--104, 2018.

\bibitem{10181313}
M.~A. Khan, M.~S. Khan, I.~Khan, S.~Ahmad, and S.~Huda, ``Non functional requirements identification and classification using transfer learning model,'' \emph{IEEE Access}, vol.~11, pp. 74\,997--75\,005, 2023.

\bibitem{HumanEvaluationDataset}
\BIBentryALTinterwordspacing
A.~Anonymous, ``Exploring large language models for automated non- functional requirements generation,'' Mar. 2025. [Online]. Available: \url{https://doi.org/10.5281/zenodo.15002723}
\BIBentrySTDinterwordspacing

\bibitem{glinz2007non}
M.~Glinz, ``On non-functional requirements,'' in \emph{15th IEEE international requirements engineering conference (RE 2007)}.\hskip 1em plus 0.5em minus 0.4em\relax IEEE, 2007, pp. 21--26.

\bibitem{Chung2009}
\BIBentryALTinterwordspacing
L.~Chung and J.~C.~S. do~Prado~Leite, \emph{On Non-Functional Requirements in Software Engineering}.\hskip 1em plus 0.5em minus 0.4em\relax Berlin, Heidelberg: Springer Berlin Heidelberg, 2009, pp. 363--379. [Online]. Available: \url{https://doi.org/10.1007/978-3-642-02463-4_19}
\BIBentrySTDinterwordspacing

\bibitem{paech2004non}
B.~Paech and D.~Kerkow, ``Non-functional requirements engineering-quality is essential,'' in \emph{10th international workshop on requirments engineering foundation for software quality}, 2004.

\bibitem{Schick2021}
T.~Schick and H.~Sch{\"u}tze, ``Exploiting cloze-questions for few-shot text classification and natural language inference,'' in \emph{Proceedings of the 16th Conference of the European Chapter of the Association for Computational Linguistics: Main Volume}.\hskip 1em plus 0.5em minus 0.4em\relax Association for Computational Linguistics, 2021.

\bibitem{ISO25010}
\BIBentryALTinterwordspacing
{International Organization for Standardization}, ``{ISO/IEC 25010:2023 - Systems and software engineering — Systems and software Quality Requirements and Evaluation (SQuaRE) — Product quality model},'' 2023, accessed: 2025-02-26. [Online]. Available: \url{https://www.iso.org/standard/78176.html}
\BIBentrySTDinterwordspacing

\bibitem{umar2024advances}
M.~A. Umar and K.~Lano, ``Advances in automated support for requirements engineering: a systematic literature review,'' \emph{Requirements Engineering}, vol.~29, no.~2, pp. 177--207, 2024.

\bibitem{cleland2007automated}
J.~Cleland-Huang, R.~Settimi, X.~Zou, and P.~Solc, ``Automated classification of non-functional requirements,'' \emph{Requirements engineering}, vol.~12, pp. 103--120, 2007.

\bibitem{casamayor2010identification}
A.~Casamayor, D.~Godoy, and M.~Campo, ``Identification of non-functional requirements in textual specifications: A semi-supervised learning approach,'' \emph{Information and Software Technology}, vol.~52, no.~4, pp. 436--445, 2010.

\bibitem{haque2019non}
M.~A. Haque, M.~A. Rahman, and M.~S. Siddik, ``Non-functional requirements classification with feature extraction and machine learning: An empirical study,'' in \emph{2019 1st international conference on advances in science, engineering and robotics technology (ICASERT)}.\hskip 1em plus 0.5em minus 0.4em\relax IEEE, 2019, pp. 1--5.

\bibitem{Rejithkumar2025NICE}
G.~Rejithkumar and P.~R. Anish, ``{NICE: Non-Functional Requirements Identification, Classification, and Explanation Using Small Language Models},'' in \emph{Proceedings of the 46th International Conference on Software Engineering: Software Engineering in Practice (ICSE-SEIP 2025)}, 2025.

\bibitem{damirchi2023non}
R.~Damirchi and A.~Amini, ``Non-functional requirement extracting methods for ai-based systems: A survey,'' in \emph{2023 13th International Conference on Computer and Knowledge Engineering (ICCKE)}.\hskip 1em plus 0.5em minus 0.4em\relax IEEE, 2023, pp. 535--539.

\bibitem{FR_NFR_Dataset}
\BIBentryALTinterwordspacing
S.~Sonali and S.~Thamada, ``{FR\_NFR\_dataset},'' 2024. [Online]. Available: \url{https://data.mendeley.com/datasets/4ysx9fyzv4/1}
\BIBentrySTDinterwordspacing

\bibitem{PURE_Dataset}
\BIBentryALTinterwordspacing
A.~Ferrari, G.~O. Spagnolo, and S.~Gnesi, ``Pure: a dataset of public requirements documents,'' sep 2018. [Online]. Available: \url{https://doi.org/10.5281/zenodo.1414117}
\BIBentrySTDinterwordspacing

\bibitem{Asif2019}
M.~Asif, M.~Malik, M.~Chaudary, E.~D.~S. Tayyaba, and M.~Mahmood, ``Annotation of software requirements specification (srs), extractions of nonfunctional requirements, and measurement of their tradeoff,'' \emph{IEEE Access}, vol.~PP, pp. 1--1, 03 2019.

\bibitem{villamizar2021requirements}
H.~Villamizar, T.~Escovedo, and M.~Kalinowski, ``Requirements engineering for machine learning: A systematic mapping study,'' in \emph{2021 47th Euromicro conference on software engineering and advanced applications (SEAA)}.\hskip 1em plus 0.5em minus 0.4em\relax IEEE, 2021, pp. 29--36.

\bibitem{de2025classification}
V.~De~Martino and F.~Palomba, ``Classification and challenges of non-functional requirements in ml-enabled systems: A systematic literature review,'' \emph{Information and Software Technology}, p. 107678, 2025.

\bibitem{habibullah2024scoping}
K.~M. Habibullah, J.~G. Diaz, G.~Gay, and J.~Horkoff, ``Scoping of non-functional requirements for machine learning systems,'' in \emph{2024 IEEE 32nd International Requirements Engineering Conference (RE)}.\hskip 1em plus 0.5em minus 0.4em\relax IEEE, 2024, pp. 496--497.

\bibitem{nascimento2021approach}
N.~Nascimento, P.~Alencar, and D.~Cowan, ``An approach to support human-in-the-loop big data software development projects,'' in \emph{2021 IEEE International Conference on Big Data (Big Data)}.\hskip 1em plus 0.5em minus 0.4em\relax IEEE, 2021, pp. 2319--2326.

\bibitem{melo2023identifying}
G.~Melo, N.~Nascimento, P.~Alencar, and D.~Cowan, ``Identifying factors that impact levels of automation in autonomous systems,'' \emph{IEEE Access}, vol.~11, pp. 56\,437--56\,452, 2023.

\bibitem{iso2022iso}
I.~ISO, ``Iso/iec 27001: 2022 information security, cybersecurity and privacy protection—information security management systems—requirements,'' \emph{International Organization for Standardization (ISO) and International Electrotechnical Commission (IEC)}, 2022.

\bibitem{ISO25058}
------, ``{ISO/IEC TS 25058:2024 - Systems and software engineering — Systems and software Quality Requirements and Evaluation (SQuaRE) — Guidance for quality evaluation of artificial intelligence (AI) systems},'' \emph{International Organization for Standardization (ISO) and International Electrotechnical Commission (IEC)}, January 2024, published by ISO/IEC JTC 1/SC 42, 20 pages.

\end{thebibliography}

\end{document}